\def\xdx{x_\lambda{\partial \over \partial x_{\lambda}}}
\def\wam{\widetilde A_-}
\def\wap{\widetilde A_+}
\def\hatk{\hat {\kappa}}

\def\rvec{{\bf r}}
\def\vec#1{\vert #1 \rangle}
\def\ksum{\sum_{\kappa=-1,2}}
\def\fun#1#2{#1^{(#2)}}

\def\dpar#1#2{{\partial #1 \over \partial #2}}

\documentstyle[12pt]{article}

\title{RELATIVISTIC DYNAMICAL POLARIZABILITY OF HYDROGEN-LIKE ATOMS} 

\author {Le Anh Thu \dag, Le Van Hoang \dag \ddag,
Komarov L I \dag , Romanova T S *\\
\dag \ \small Department of Theoretical Physics,
Belarussian State University\\
\small 4  Fr.Skariny av., Minsk 220050, Republic of Belarus\\ 
\ddag \ \small Institute of Physico-Chemical Problems,
Belarussian State University\\
\small 14 Leningradskaya str., Minsk 220080, Republic of Belarus\\
{}*\ \small 
Institute of Nuclear Problems, Belarussian State University\\}

\date{}

\begin{document}

\maketitle
\noindent

%%The short title\\
%%{RELATIVISTIC DYNAMICAL POLARIZABILITY }
%%\vskip 2cm
%%\noindent
%%Classification number
%%\begin{description}
%%\item[00.00] GENERAL
%%\item[03.00] Classical and Quantum Physics; Mechanics and Field
%%\item[03.65] Quantum Theory; Quantum Mechanics
%%\end{description}
\begin{description}
\item[PACS.32.80Wr]-Other multiphoton processes.
\item[PACS.03.65Pm]-Relativistic wave equations.
\item[PACS.31.15Md]-Perturbation theory.
\end{description}
\vskip 2cm
%%\centerline {\bf Abstract}
\begin{abstract}
Using the operator representation of the Dirac Coulomb Green function
the analytical method in perturbation theory is employed in obtaining 
solutions of the Dirac equation for a hydrogen-like atom in a time-dependent 
electric field. The relativistic dynamical polarizability of hydrogen-like 
atoms is calculated and analysed.
\end{abstract}
\newpage

\section{Introduction}

In stationary perturbation theory as well as in the time-dependent one, the method 
of using the Coulomb Green function has found wide application in obtaining 
analytical solutions of the Schr\"odinger or Dirac equations. The main
advantage of this method is the possibility of obtaining the final results 
in closed analytical form or reducing results to the summation of rapidly
convergent series. Therefore, by using the above-mentioned method one
can avoid
a lot of complex numerical integrations (see, for example, Makhan\"ek and
Korol'kov 1982, Zapryagaev {\it et al} 1985 and references cited therein).
On the basis of the application of the connection between the problem of the 
four-dimensional isotropic 
harmonic oscillator and that of a hydrogen-like atom in eletromagnetic fields
(see Kustaanheimo and Stiefel 1965), it has been proposed to establish a new
representation of the Coulomb Green function in the form of a product of annihilation and
creation operators (this is called by us an operator representation). Operator
representation of the Coulomb Green function is very efficient in application for
the non-relativistic case (Le Van Hoang {\it et al} 1989) as well as
for the relativistic one (Le Anh Thu {\it et al} 1994). The main elements of the
algebraic method used (with the aid of the operator representation of the Coulomb Green
functions) in Le Van Hoang {\it et al} (1989) and in Le Anh Thu {\it et al} (1994) are as
follows. By using the above-mentioned connection, all operators of the algebra of the
dynamical symmetry group $SO(4,2)$ can be found in the quadratic form of the annihilation
and creation operators (see, for example, Kleinert 1968, Komarov and Romanova 1982).
Therefore, the calculation method, based on the use of the algebra of $SO(4,2)$, leads
only to the use of the simple communtaion relations between the latter operators. The use 
of this method
together with the operator
representation of the Coulomb Green function essentially reduces the calculation
process and provides for reducing rather complicated calculations of matrix
elements with the Coulomb wavefunctions to a purely algebraic procedure of
transforming the product of the annihilation and creation operators to the
normal form. The advantage of the proposed algebraic method are found not only in the
simplicity of the calculation process but also in the possibility of obtaining the final
results in the summation of rapidly convergent series. In fact, some clever results are
obtained in Le Anh Thu {\it et al} (1994) for the problem of calculation of the
relativistic polarizability of hydrogen-like atoms. In this present paper, we consider the problem of calculating the 
relativistic dynamical polarizability of the ground state of hydrogen-like 
atoms on the basis of application of the operator representation of
the Dirac Coulomb Green 
function established in Le Anh Thu {\it et al} 1994. These calculations,
besides their  
purely theoretical significance, are of great practical interest
connected with  
recent developments in  experimental investigations of multiply charged 
ions (see, for example, Zapryagaev {\it et al} 1985, Paratzacos and
Mork 1979). However, the majority of accurate calculations has been done only for the
static polarizability of relativistic hydrogen-like atoms (see also Drake and Goldman
1981, 1988, Johnson {\it et al} 1988). In our calculations 
the radiation corrections are neglected, taking into account the fact
that this 
effect is small compared with the external field effect. Our results are 
directly generalized from the non-relativistic calculations 
(Zapryagaev {\it et al} 1985) and coincide with the results in the static
limit (Le Anh Thu {\it et al} 1994).

\section{Equation in two-dimensional complex space}

The Dirac equation for a hydrogen-like atom in the field of linearly 
polarized light can be written as follows ($\hbar = m = c= 1$):

  $$ \left(-i\alpha_{\lambda}\; r {\partial \over \partial x_{\lambda}} + 
     \beta r - Ze^2  + {\theta \over 2}\; erx_3 \left(e^{i\nu t} +
     e^{-i\nu t}\right) \right )\Psi (\rvec,t)= $$

   $$ =ir\,\dpar{\Psi (\rvec,t)}{t}, 
   \eqno(1) $$
where $\alpha_{\lambda}(\lambda=1,2,3)$ and $\beta$ are the Dirac
matrices ; $\theta$ and $\nu$ are the amplitude and 
frequency of the external electric field respectively. Further on, we use the usual 
representation

 $$ \Psi = \left(\matrix{\Psi_1\cr \Psi_2 \cr}\right), \qquad
    \alpha_{\lambda} = \left(\matrix{0 & \sigma_{\lambda}\cr
    \sigma_{\lambda} & 0 \cr}\right), \qquad
    \beta = \left(\matrix{1 & 0 \cr 0 & -1\cr}\right), $$
where $\Psi_1$ and $\Psi_2$ are two-component spinors and $\sigma_{\lambda}\;
(\lambda=1,2,3)$ are the Pauli matrices.

The formal changes  (see Le Anh Thu {\it et al} 1994)

$$ x_{\lambda}\rightarrow (\sigma_{\lambda})_{st}\xi^*_s\xi_t\;,
    \qquad r \rightarrow \xi^*_s\xi_s\;,$$

$$ r\hat p_{\lambda} \rightarrow -{i \over 2}(\sigma_{\lambda})_{st}
     \left( \xi_t{\partial \over \partial \xi_s} +  
     \xi^*_s{\partial \over \partial \xi^*_t} \right)\eqno(2)$$
reduce equation (1) to an equation describing the interaction between a 
``particle'' with complex coordinates $\xi_s \; (s= 1, 2)$ and the
external electric field. Here, in (2) summation is indicated
by means of repeated indices. The scalar product of wavefunctions in 
$\xi$--space is defined by the relation 

$$\left\langle \widetilde \Psi \vert \widetilde \varphi \right\rangle =
   \int_{-\infty}^{+\infty}d\,\xi_{1}^{'} \int_{-\infty}^{+\infty}d\,
   \xi_{1}^{''} \int_{-\infty}^{+\infty}d\,\xi_{2}^{'} 
   \int_{-\infty}^{+\infty}d\,\xi_{2}^{''} \; \widetilde 
   \Psi^*(\xi_1',\xi_1'',\xi_2',\xi_2'')
   \; \widetilde \varphi (\xi_1',\xi_1'',\xi_2',\xi_2''), 
   \eqno(3) $$
where $\xi_s' \equiv Re\xi_s$, $\xi_s'' \equiv Im\xi_s $.
All operators appearing in (1) are henceforth considered to conform 
with the formal changes (2). Thus, the operator in the left-hand side
of equation (1) is self-adjoint with respect to the scalar product of 
wavefunctions defined by (3).

We will solve equation (1) by using the method of perturbation theory 
assuming the external electric field to be small. Its solution can be 
found in the form 

  $$ \Psi (\rvec,t)=\Psi^{(0)} (\rvec,t) + \theta \Psi^{(1)} (\rvec,t)$$
  $$\qquad \qquad\equiv \Psi^{(0)} (\rvec)\;e^{-i\varepsilon_0 t} + \theta 
    u(\rvec)\;e^{-it(\varepsilon_0 -\nu)} + \theta v(\rvec)
    \;e^{-it(\varepsilon_0 +\nu)},
   \eqno(4)$$
where $\Psi^{(0)} (\rvec) $ is a wavefunction in the zero-order 
approximation, i.e. a solution of the Dirac equation

  $$\left(-i\alpha_{\lambda}\; r {\partial \over \partial x_{\lambda}} + 
     \beta r -\varepsilon_0 r \right) \Psi^{(0)} (\rvec)
    =Ze^2\;\Psi^{(0)} (\rvec) \; ;\eqno(5)$$
$\varepsilon_0 $ is the energy in the zero-order
approximation. By substituting (4) into (1) and taking into account (5),
we then obtain the equations

  $$\left(-i\alpha_{\lambda}\; r {\partial \over \partial x_{\lambda}} + 
     \beta r - Ze^2 \right)u(\rvec)-r(\varepsilon_0-\nu)\;u(\rvec)
     =-{1 \over 2}\;erx_3 \Psi^{(0)} (\rvec) \eqno(6)$$

  $$\left(-i\alpha_{\lambda}\; r {\partial \over \partial x_{\lambda}} + 
     \beta r - Ze^2 \right)v(\rvec)-r(\varepsilon_0+\nu)\;v(\rvec)
     =-{1 \over 2}\;erx_3 \Psi^{(0)} (\rvec). \eqno(7)$$
Noting that equations (6) and (7) have the same structure we consider
only equation (6) for the function $u({\bf r})$; then by replacing
$-\nu$ by $\nu$ in $u({\bf r})$ we find the function $v({\bf r})$.

Let us now present $u({\bf r})$ and $\Psi^{(0)}$ in the form 

  $$u= {\varphi_1^{(1)}\choose ( \sigma_{\lambda} x_{\lambda}/ r)
     \;\varphi_2^{(1)}},  \qquad
  \Psi^{(0)} ={\varphi_1^{(0)}\choose (\sigma_{\lambda} x_{\lambda}/ r)
   \;\varphi_2^{(0)}}.  \eqno(8) $$
The substitution of (8) into (6) leads to the set of equations for
$\fun{\varphi_1}{1}$ and $\fun{\varphi_2}{1}$ :

  $$-i\left(\xdx+1\right)\fun{\varphi_2}{1}-i\hatk\fun{\varphi_2}{1} +
     r(1-\varepsilon_0+\nu)\fun{\varphi_1}{1}-Ze^2 \fun{\varphi_1}{1}
     =-{1 \over 2}erx_3\fun{\varphi_1}{0} \eqno(9)$$

   $$-i\left(\xdx+1\right)\fun{\varphi_1}{1}+i\hatk\fun{\varphi_1}{1} -
     r(1+\varepsilon_0-\nu)\fun{\varphi_2}{1}-Ze^2 \fun{\varphi_2}{1}
     =-{1 \over 2}erx_3\fun{\varphi_2}{0}, \eqno(10)$$	
where $\hatk =1+\sigma_{\lambda}\hat l_{\lambda}$ ; $\hat{\bf l}$  
is the orbital momentum operator.

By using the transformations

     $$ \fun{\varphi_1}{0}=-{i \over 2}\sqrt{1+\varepsilon_0}
      \left({\cal F}^{(0)} -{\cal G}^{(0)}\right), \qquad 
      \fun{\varphi_2}{0}={1 \over 2}\sqrt{1-\varepsilon_0}
      \left({\cal F}^{(0)} +{\cal G}^{(0)}\right) \eqno(11.1) $$
and
     $$ \fun{\varphi_1}{1}=-{i \over 2}\sqrt{1+\varepsilon_0-\nu}
      \left({\cal F}^{(1)} -{\cal G}^{(1)}\right), \qquad
      \fun{\varphi_2}{1}={1 \over 2}\sqrt{1-\varepsilon_0+\nu}
      \left({\cal F}^{(1)} +{\cal G}^{(1)}\right) \eqno(11.2)$$	
we find	$$ \left[\xdx+1+\omega r-{Ze^2 \over \omega}(\varepsilon_0-\nu)\right]
     {\cal F}^{(1)}+\left(\hatk+{Ze^2 \over \omega}\right){\cal G}^{(1)}$$

   $$ \qquad = rx_3(\wam {\cal F}^{(0)}+\wap {\cal G}^{(0)}), \eqno(12)$$

   $$ \left[\xdx+1-\omega r+{Ze^2 \over \omega}(\varepsilon_0-\nu)\right]
      {\cal G}^{(1)}+\left(\hatk-{Ze^2 \over \omega}\right){\cal F}^{(1)}$$ 
   $$ \qquad = -rx_3(\wap {\cal F}^{(0)}+\wam {\cal G}^{(0)}), \eqno(13)$$
where   

   $$\omega=\sqrt{1-(\varepsilon_0-\nu)^2},$$
   $$\widetilde A_{\pm}={e \over 4\omega}\left[\sqrt{(1-\varepsilon_0)
     (1-\varepsilon_0+\nu)}\pm\sqrt{(1+\varepsilon_0)
     (1+\varepsilon_0-\nu)}\right].\eqno(14)$$                
After expanding the functions ${\cal F}$ and ${\cal G} $ in power
series of the eigenfunctions of operators $\hat L^2$ and $\hatk$, the 
operator $\hatk$ appearing in (12) and  (13) becomes a c-number and, 
therefore, in order to solve equations (12) and (13) we can employ 
the Green function method established in Le Anh Thu {\it et al} (1994).
In the next section we will show an example by calculating the
dynamical polarizability of hydrogen-like atoms in the ground state.

\section{Relativistic dynamical polarizability in the ground state 
	of hydrogen-like atoms}

Solution of equation (5) can be easily obtained by purely algebraic
calculations. In particular, we can find for the
ground state the solution (see Komarov and Romanova 1985):

  $$ {\cal F}^{(0)}={(2\omega_0)^{\varepsilon_0-1} \over 
     \sqrt{\Gamma(2\varepsilon_0)}}r^{\varepsilon_0-1}\vec{0(\omega_0)}
     \chi^{\uparrow},$$

  $$ {\cal G}^{(0)}=0,\eqno(15)$$
where $\omega_0=Ze^2$ ; $\varepsilon_0=\sqrt{1-(Ze^2)^2}$; 
$\chi^{\uparrow(\downarrow)}$ are eigenvectors of operator $\sigma_3$ and 
$\vec{0(\omega_0)}$ is the vacuum state defined by the equations
 $$a_s(\omega_0)\vec{0(\omega_0)}=b_s(\omega_0)\vec{0(\omega_0)}=0, 
     \qquad s=1,2. \; $$
Here the operators $a_s(\omega)$, $b_s(\omega)$ are defined as follows

$$  a_s(\omega)=\sqrt{\frac{\omega}{2}}\left(\xi_s+\frac{1}{\omega}
       \frac {\partial}{\partial \xi^*_s}\right), \qquad  
       b_s(\omega)=\sqrt{\frac{\omega}{2}} \left(\xi^*_s+\frac{1} {\omega}\frac
       {\partial}{\partial \xi_s}\right),$$ 

    $$  a^+_s(\omega)=\sqrt{\frac{\omega}{2}}\left(\xi^*_s-\frac{1}{\omega}
        \frac {\partial}{\partial \xi_s}\right), \qquad
       b^+_s(\omega)=\sqrt{\frac{\omega}{2}} \left(\xi_s - \frac{1} 
       {\omega}\frac {\partial}{\partial\xi^*_s}\right)$$
where $\omega$ is a positive parameter (see Le Anh Thu {\it et al} 1994).
The perturbation term in equations (12) and (13) thus has the form

  $$ \pm\widetilde A_{\mp}rx_3{\cal F}^{(0)}=\pm A_{\mp}r^{\varepsilon_0}
      (Z_{1,-1}+\sqrt 2 Z_{1,2})\vec{0(\omega_0)}.\eqno(16)$$
Here, we use the notation

  $$  A_{\pm}={\sqrt 6(2\omega_0)^{\varepsilon_0-2}\over 
       \sqrt{\Gamma(2\varepsilon_0)}}\widetilde A_{\pm},$$
and $Z_{l,\kappa}$ are eigenvectors of (i) the square orbital
momentum operator and (ii) the operator $\hatk$.
The structure of perturbation term (16) prompts us to find the solution
in the form

  $$ {\cal F}^{(1)}=\ksum {\cal F}^{(1)}_{\kappa}, \qquad
    {\cal G}^{(1)}=\ksum {\cal G}^{(1)}_{\kappa} \eqno(17)$$
where ${\cal F}^{(1)}_{\kappa}$ and ${\cal G}^{(1)}_{\kappa}$ satisfy 
the equations

  $$ \left[\xdx+1+\omega r-{Ze^2 \over \omega}(\varepsilon_0-\nu)\right]
     {\cal F}^{(1)}_{\kappa}+\left(\kappa+{Ze^2 \over \omega}\right)
     {\cal G}^{(1)}_{\kappa}$$
   $$\qquad = A_-\sqrt{\vert \kappa\vert}r^{\varepsilon_0}Z_{1\kappa}
     \vec{0(\omega_0)}, \eqno(18)$$
   $$ \left[\xdx+1-\omega r+{Ze^2 \over \omega}(\varepsilon_0-\nu)\right]
     {\cal G}^{(1)}_{\kappa}+\left(\kappa-{Ze^2 \over \omega}\right)
     {\cal F}^{(1)}_{\kappa}$$
   $$\qquad = -A_+\sqrt{\vert \kappa\vert}r^{\varepsilon_0}Z_{1\kappa}
     \vec{0(\omega_0)}. \eqno(19)$$
From (19) it follows that

    $${\cal F}^{(1)}_{\kappa}=-{A_+ \sqrt{\vert \kappa\vert} \over 
     \kappa-{Ze^2\over\omega}}r^{\varepsilon}Z_{1\kappa} 
    \vec{0(\omega_0)}-{1\over \kappa-{Ze^2\over\omega}}
     \biggl[\xdx+1-\omega r+{Ze^2 \over \omega}(\varepsilon_0-\nu)\biggr]
     {\cal G}^{(1)}_{\kappa}.\eqno(20)$$
Substituting (20) into (18) we obtain
$$\Biggl\{-\left[\xdx+1+\omega r-{Ze^2 \over \omega}(\varepsilon_0-\nu)\right] 
   \left[\xdx+1-\omega r+{Ze^2 \over \omega}(\varepsilon_0-\nu)\right] $$

$$ +\kappa^2-\biggl({Ze^2 \over \omega}\biggr)^2 \Biggr\}
   {\cal G}^{(1)}_{\kappa}=
    A_-\sqrt{\vert \kappa\vert}\biggl(\kappa-{Ze^2 \over \omega}\biggr)
    r^{\varepsilon_0}Z_{1\kappa}\vec{0(\omega_0)} $$
 $$ + A_+\sqrt{\vert \kappa\vert}
   \left[\xdx+1+\omega r-{Ze^2 \over \omega}(\varepsilon_0-\nu)\right] 
   r^{\varepsilon_0}Z_{1\kappa}\vec{0(\omega_0)}. \eqno(21)$$	
As mentioned above (see (2)), all operators in (20), (21)  (though formally written for
brevity via the usual coordinates $x_{\lambda}$ ($\lambda=1,2,3$)) are understood in the
sence of the formal changes (2). These changes are equivalent to the transformation from
$r$-space to $\xi$-space (Kustaanheimo-Stiefel transformation, see Kustaanheimo-Stiefel
(1965)):

$$\cases{ x_{\lambda} = \xi^*_s (\sigma_{\lambda})_{st} \xi_t \qquad (s,t=1,2)\cr
               \chi     = \arg (\xi_1)\cr} $$
So we can rewrite the operators used in (20), (21) via the annihilation and creation
operators as follows 
  $$ \xdx={M(\omega)\over 2}-{M^+(\omega)\over 2}-1,$$

   $$r={1\over 2\omega}[M(\omega)+M^+(\omega)+N(\omega)+2]$$
where the notation
$$N=a_s^+a_s+b_s^+b_s \qquad M=a_sb_s \qquad M^+=a^+_sb^+_s.$$
Therefore, equation (21) has a more convenient form
$$\Biggl\{\biggl[M+{N\over 2}+1-{Ze^2 \over \omega}(\varepsilon_0-\nu)\biggr]
 \biggl[M^++{N\over 2}+1-{Ze^2 \over \omega}(\varepsilon_0-\nu)\biggr]$$ 

  $$ +\kappa^2-\biggl({Ze^2 \over \omega}\biggr)^2\Biggr\}
      \vec{{\cal G}^{(1)}_{\kappa}}=
    A_-\sqrt{\vert \kappa\vert}\biggl(\kappa-{Ze^2 \over \omega}\biggr)
    r^{\varepsilon_0}Z_{1\kappa}\vec{0(\omega_0)}$$
$$ + A_+\sqrt{\vert \kappa\vert}
    \biggl[M+{N\over 2}+1-{Ze^2 \over \omega}(\varepsilon_0-\nu)\biggr]
      r^{\varepsilon_0}Z_{1\kappa}\vec{0(\omega_0)}. \eqno(22) $$
Here and henceforth, we omit for brevity the parameter $\omega$ in the 
expressions of the operators. Equation (22) has the same structure as the equations
appearing in the case of calculation of the static polarizability of hydrogen-like atoms
(see  Le Anh Thu {\it et al} 1994). The only difference is in the perturbation term on
the left-hand side of these equations. Therefore, we can by analogy find the solutions of
equation (22), using the Green function operator which, 
according to Le Anh Thu {\it et al} (1994), can be
established as follows:

 $$\hat{\bf G}_{l\kappa}=\hat B_{l\kappa} \frac{1}{2+N+M+M^+},\qquad 
   \hat B_{l\kappa}=\sum_{s=0}^{\infty} d_s(N/2)M^s ,\eqno(23) $$
where

   $$ d_o(n) = -\frac{1}{n+\gamma+2-Ze^2\varepsilon/2} ,\eqno(24)$$
  
    $$ d_s(n)=(-1)^s(2\gamma+1)\frac{(n+l+1)!}{(n+s+l+1)!}{\Gamma
     (n+s+1-\gamma-Ze^2\varepsilon/2)\over \Gamma(n+2-\gamma-Ze^2\varepsilon
       /\omega)} $$
   $$ \qquad\times{\Gamma(n+2+\gamma- Ze^2\varepsilon/2)\over
    \Gamma(n+s+3+\gamma-Ze^2\varepsilon/2)},\quad  s= 1,2,3\ldots $$
  $$\gamma=-l-1+\sqrt{\kappa^2-Z^2e^4}, \quad \varepsilon
   =\varepsilon_0-\nu.$$				
Here, we note that the Green function operator (23) acts on the basis
of states with frequency $\omega$. However, the wavefunctions in the zero-order 
approximation (15) have the frequency $\omega_0 = Ze^2$. Therefore, in order 
to use the algebraic calculation we have to transform the wavefunctions (15) 
from the frequency $\omega_0$ to $\omega$ using the 
unitary transformation (see Komarov and Romanova 1982):

$$\vec{\Phi(\omega_0)}=\hat U(\omega_0,\omega)\vec{\Phi(\omega)}, $$
where 

$$\hat U(\omega_0,\omega)=\exp\left\{{1\over 2}\ln{\omega_0\over\omega}
    [M(\omega)-M^+(\omega)]\right\}.$$
This transformation can be reduced to the normal form

  $$ \hat U(\omega_0,\omega)={4\omega_0\omega\over (\omega_0+\omega)^2}
     \exp\Biggl({\omega-\omega_0 \over \omega+\omega_0}M^+(\omega)\Biggr)
 \exp\Biggl(N(\omega)\ln{2\sqrt{\omega\omega_0}\over\omega+\omega_0}\Biggr)$$ 

  $$\times\exp\Biggl(-{\omega-\omega_0 \over \omega+\omega_0}M(\omega)\Biggr).
   \eqno(25)$$	
Finally, by analogy with that was done in Le Anh Thu {\it et al} (1994)
for the same equation, we find the solution

   $$ \vec{{\cal G}^{(1)}_{\kappa}}=
    A_-\sqrt{\vert \kappa\vert}\Bigl(\kappa-{Ze^2 \over \omega}\Bigr)
    r^{\gamma}\hat{\bf G}_{1\kappa}r^{\varepsilon_0-\gamma}
    \hat U(\omega_0,\omega)Z_{1\kappa}\vec{0(\omega_0)}+$$

   $$+A_+\sqrt{\vert \kappa\vert}r^{\gamma}\hat{\bf G}_{1\kappa}r^{-\gamma}
     \biggl[M+{N\over 2}+1-{Ze^2 \over \omega}(\varepsilon_0-\nu)\biggr]
    r^{\varepsilon_0} \hat U(\omega_0,\omega)Z_{1\kappa}\vec{0(\omega_0)}.
   \eqno(26)$$	
The wavefunction $\vec{{\cal F}^{(1)}_{\kappa}}$ can be obtained by 
substituting (26) into (20).

Let us now calculate the dynamical polarizability of the ground state 
of hydrogen-like atoms, the formula of which in $\xi$-space has the form

   $$ a(\nu)=2{\langle\Psi^{(0)}\vert erx_3\vec{\Psi^{(1)}}\over
     \langle\Psi^{(0)}\vert r \vec{\Psi^{(0)}}} $$
This formula, after considering (8), (11), (15) and (17), can be written as
follows 

      $$ a(\nu)=\ksum a_{\kappa}(\nu), $$
where  

   $$ a_{\kappa}(\nu)={4\omega_0\omega\sqrt{\vert \kappa \vert} \over
    \varepsilon_0}\langle 0(\omega_0)\vert Z_{1\kappa}r^{\varepsilon_0}
    \vert A_+{\cal F}^{(1)}_{\kappa}+A_-{\cal G}^{(1)}_{\kappa}\rangle.
   \eqno(27) $$ 
By substituting the found solutions (20), (26) into (27) we find the 
expression for the positive frequency term of polarizability 

   $$ a_{\kappa}(+\nu)={4\omega_0\omega\vert \kappa\vert \over \varepsilon_0}
      \Biggl\{{A_+^2\over B}\langle 0 \vert 
      r^{2\varepsilon_0} \vec 0 
    +\Biggl [A_-^2 B+ 2A_{+}A_{-}\delta+{A_+^2\delta^2
    \over B}\Biggr] H^{\kappa}_{00}$$

  $$ +{4\omega(\omega-\omega_0) \over (\omega+\omega_0)^2}
   \Biggl[A_+A_- +{A_+^2 \delta \over B}\Biggr]
   H^{\kappa}_{01} + {4\omega^2(\omega-\omega_0)^2 \over (\omega+\omega_0)^4}
   {A_+^2\over B} H^{\kappa}_{11}\Biggr\}. \eqno(28)$$
Here, we use the notations
  $$ B = \kappa-{Ze^2\over\omega}, $$
  $$ \delta={4(\omega-\omega_0) \over \omega+\omega_0}-
     {Ze^2 \over \omega}(\varepsilon_0-\nu)
     +\varepsilon_0+2, \eqno(29)$$

  $$ H^{\kappa}_{nm}=\langle 0(\omega) \vert Z_{1\kappa}\hat U^{+}
     (\omega_0,\omega)
    M^n\,r^{\varepsilon_0+\gamma}\hat {\bf G}_{\kappa} 
    r^{\varepsilon_0-\gamma}(M^+)^m \hat U(\omega_0,\omega) 
   Z_{1\kappa}\vec {0(\omega)}.$$
A similar expression for the negative frequency term can be obtained by 
replacing $\nu$ by $-\nu$ in formulae (14), (28), (29).

By using the algebra of operators $M^+,\,M,\,N$:

   $$ \left [M, M^{+}\right] = N + 2,\qquad \left [M, N+2\right] = 2 M,
   \qquad \left [N+2, M^{+}\right] = 2 M^{+} $$
and the relations

  $$N(M^+)^n Z_{l\kappa}\vec 0=2(n+l)(M^+)^n Z_{l\kappa}\vec 0 ,$$

  $$M(M^+)^n Z_{l\kappa}\vec 0=n(n+2l+1)(M^+)^{n-1} Z_{l\kappa}\vec 0 ,$$

  $$r^{\rho} Z_{l\kappa}\vec 0={\Gamma(\rho+2l+2)\over (2\omega)^{\rho}
    \Gamma(-\rho)}\sum_{s=0}^{\infty}{(-1)^s \over s!}{\Gamma(s-\rho)
    \over (s+2l+1)!} (M^+)^s Z_{l\kappa} \vec 0 ,$$
we algebraically obtained the explicit form of the 
term $H^{\kappa}_{nm}$ as follows:

  $$H^{\kappa}_{nm}={1\over 12\omega}\biggl( {2\sqrt{\omega_0\omega} 
    \over \omega_0+\omega} \biggr)^8 (\omega_0+\omega)^{1-2\varepsilon_0}
    \sum_{p=0}^{\infty}(p+3)! \sum_{q=0}^p(-1)^q {p \choose q}$$

  $$\times {\Gamma(q+\varepsilon_0-\gamma)\,\Gamma(q+3+\varepsilon_0-\gamma)
    \over \Gamma(q+\varepsilon_0-\gamma-m)\,(q+3)!}\sum_{s=0}^p
    (-1)^s {d_s(p-s+1) \over (p-s)!} \sum_{t=0}^{p-s}(-1)^t {p-s \choose t}$$

   $$\times\biggl({2\omega \over \omega+\omega_0}\biggr)^{t+q-n-m}
     {\Gamma(t+1+\varepsilon_0+\gamma)\,\Gamma(t+4+\varepsilon_0+\gamma)
    \over \Gamma(t+1+\varepsilon_0+\gamma-n)\,(t+3)!}.\eqno(30)$$
Direct calculations show that all power series appearing in the term 
$H^{\kappa}_{nm}$ are 
rapidly convergent. The high convergency of these power series is directly 
related to the expansion (23) of the Dirac Coulomb Green function which, 
in fact, is established on the basis of harmonic oscillator wavefunctions. 
Moreover, the expression (28) for the polarizability contains 
only $H^{\kappa}_{nm}$ with 
$n, m =0, 1$, the calculation of which needs only some first terms in the 
summation over $p$. For example, for frequency $\nu$ less than the 
fourth resonance frequency 
(relative to the transition from ground state to $3P_{3/2}$ state) the 
contribution of the terms with $p=0, 1$ in $H^{\kappa}_{nm}\; (n, m=0, 1)$ 
is about $98\;\%$--$99\;\%$ for all $Z \leq 137$. 

In figures 1(a)and  1(b) the dependence of relativistic 
polarizability on the external field frequency is given for hydrogen-like 
atoms with $Z=50$. The dotted lines correspond to the non-relativistic 
limit case. Figure 2 give the same for $Z=100$.

Let us now consider the non-relativistic limit, i.e. take into account only 
the first term in the expansion in the power series of $Ze^2 $. For this case,
we can effect summation over the variables $t, q, s$ in the 
expression of $H^{\kappa}_{nm}$ and then have $H^{\kappa}_{nm}$ in the 
form of hypergeometric functions. Consequently, we have

  $$a_{non}(\pm\nu)={2^{11}\mu^3 e^2 \over (Ze^2)^4(\mu+1)^{10}} 
     \Biggl \{ {1 \over (2-{1\over\mu})(3-{1\over\mu})}\; {}_2F_1
     \Biggl(5, 2-{1\over\mu}, 4-{1\over\mu}, \biggl({\mu-1 \over \mu+1}
     \biggr)^2\Biggr)$$

   $$ + {5\mu^2 \over (\mu+1)^2(3-{1\over\mu})}\;{}_2F_1 
      \Biggl(6, 3-{1\over\mu}, 4-{1\over\mu}, \biggl({\mu-1 \over \mu+1}
     \biggr)^2\Biggr) \Biggr\},$$
where $\mu=\sqrt{1 \pm 2\nu/(Ze^2)^2}$. This expression coincides with 
the well known result obtained by Vetchinkin and Khristenko (1968).

By putting $\nu = 0$ into (28), (30) and (31) we thus find the formula 
for the relativistic polarizability. It is easy to see that for 
$\omega = \omega_0$ the formula (28) contains only the term 
$H^{\kappa}_{00}$. Taking into 
account the formula (Prudnikov {\it et al} 1981)

$$\sum_{k=0}^n(-1)^k{n \choose k} {a+k \choose m}=(-1)^n{a \choose m-n} $$
we can lead $H^{\kappa}_{00}$ to the form

  $$H^{\kappa}_{00}={1 \over 6(2\omega_0)^{2\varepsilon_0}}\>
    \Gamma(\varepsilon_0-\gamma)\>\Gamma(\varepsilon_0-\gamma+3)\>
    \Gamma(\varepsilon_0+\gamma+1)\>\Gamma(\varepsilon_0+\gamma+4)$$

   $$\times \sum_{p=0}^{\infty}{1 \over \Gamma(\varepsilon_0-\gamma-p)} 
    \sum_{s=0}^p {d_s(p-s+1) \over (p-s)!\,(p-s+3)!\, 
    \Gamma(\varepsilon_0+\gamma+1-p+s)}. \eqno(32)$$
By substituting (32) into (28) we obtain for $\nu = 0$ the  relativistic 
static polarizability

 $$ a = {e^2 (\varepsilon_0 + 1) (2\varepsilon_0 +1) (4\varepsilon_0^2
   + 13\varepsilon_0 + 12) \over 36\omega_0^4} - {e^2 (\varepsilon_0-2)^2
   \Gamma(\varepsilon_0 +\gamma +4)\,\Gamma(\varepsilon_0-\gamma+3) \over 
   36 \omega_0^4 \Gamma(2\varepsilon_0) \Gamma(-\varepsilon_0-\gamma)
   \Gamma(1-\varepsilon_0+\gamma)}$$

 $$\times \Biggl[ \sum_{k=0}^{\infty} {\Gamma(k-\varepsilon_0-\gamma)
     \Gamma(k-\varepsilon_0+\gamma+1) \over k!\, (k+3)!\,
     (k-\varepsilon_0+\gamma+3)}-(2\gamma+1) \sum_{q=1}^{\infty}
    {\Gamma(q-\varepsilon_0+\gamma+1) \over (q+3)!}$$

  $$\times {\Gamma(q-\varepsilon_0-\gamma+2) \over 
     \Gamma(q-\varepsilon_0+\gamma+4)}
    \sum_{s=0}^{q-1} {\Gamma(s-\varepsilon_0-\gamma)\,
    \Gamma(s-\varepsilon_0+\gamma+3) \over s!\, 
     \Gamma((s-\varepsilon_0-\gamma+3)} \Biggr] ,$$
where $\gamma=-2+\sqrt{4-(Ze^2)^2}$. This result absolutely coincides with
the result obtained in Le Anh Thu {\it et al} (1994) (see also Barut
and Nagel 1976).

\section{Conclusion}

In conclusion we would like to note that the magnetic field effects, as
a rule, 
should be taken into account for a detailed investigation of the behavior of 
a relativistic atom in the field of linearly polarized light. These effects can 
be neglected only in the non-relativistic limit. The above method proposed 
with the use of the operator representation of the Coulomb Green function 
can also be  employed, for example, in calculating magnetic polarizability.
Consideration of magnetic field effects leads only to an enormous number 
of calculations, which are more complicated in comparison with the above 
calculations but could be done analogously by analytical methods. Moreover, we hope that 
our algebraic method would be helpful when considering the problem of an 
atom in a quantum field, in particular in calculating the Lamb shift of 
multiply charged ions, which is of great interest and has been widely investigated 
recently (see, for example, Snyderman 1991).

\section*{Acknowledgement}
The authors would like to thank Professor A O Barut for useful
discussions and for his interest in this work.
One of the authors (LVH) would like to thank the Fundamental Research 
Foundation of the Republic of Belarus for the support rendered.

\end{document}